\documentstyle{europhys}

\def\stars{\bigskip\centerline{***}\medskip}

\newif\ifboo \boofalse

\input epsf
\begin{document}
\euro{0}{0}{1-$\infty$}{0000} \Date{} \shorttitle{Aging in spin
glass and ferromagnet}
\title{Aging phenomena in  spin glass and ferromagnetic
phases: domain growth and wall dynamics}
\author{ E. Vincent, V. Dupuis, M. Alba, J. Hammann, and
J.-P. Bouchaud} \institute{Service de Physique de l'Etat
Condens\'e, CEA Saclay, 91191 Gif sur Yvette Cedex, France
     }
\rec{1999}{in final form 2000} \pacs{ \Pacs{02}{75.50.Lk}{  Spin
glasses and other random magnets} \Pacs{03}{75.10.Nr}{  Spin-glass
and other random models} \Pacs{04}{75.40.Gb}{  Dynamic properties
(dynamic susceptibility, spin
      waves, spin diffusion, dynamic scaling, etc.)}
      }
\maketitle
\begin{abstract}
We compare aging in a disordered ferromagnet and in a spin glass, by
studying the different phases of a reentrant system. We have measured
the relaxation of the low-frequency {\it ac} susceptibility $\chi$, in
both the {\it ferromagnetic} and {\it spin-glass} phases of a
$CdCr_{1.9}In_{0.1}S_4$ sample. A restart of aging processes when the
temperature is lowered (`chaos-like' effect) is observed in both
phases. The memory of previous aging at a higher temperature can be
retrieved upon re-heating, but in the ferromagnetic phase it can
rapidly be erased by the growth of ferromagnetic domains. We interpret
the behaviour observed in the ferromagnetic phase in terms of a
combination of {\it domain growth} and {\it pinned wall reconformations},
and suggest that aging in spin glasses is dominated by such wall
reconformation processes.  

\end{abstract}
\section{Introduction}

In a spin glass, cooling below $T_g$ is the starting point of the slow
`aging' evolution. The {\it ac} susceptibility relaxes downwards with
time after the initial quench (the `age') (see e.g. \cite{Sitges} and
references therein). A natural basis for the interpretation of aging
is the idea of a slow domain growth of a spin-glass type `ordered
phase', as was developed in \cite{domains,Review}. However, several
non-trivial experimental features of aging in spin glasses are hard to
interpret within a simple domain growth scenario
\cite{hierarki,DTupps,memchaos}.  Namely, after aging at a given
temperature, aging is {\it reinitialized} by a further temperature
decrease
\footnote{This has been called `chaos', by analogy with the
theoretical scenario of \cite{bray}. However, as we discuss below,
this effect can be related to the temperature variation of Boltzmann
weights, and a better word might be `rejuvenation'.}, and the {\it
memory} of previous aging can be retrieved upon heating back.

On the other hand, domain growth phenomena should be more directly
relevant to the case of ferromagnets.  Our motivation in the present
work is to compare aging in these systems and in spin glasses, having
in mind the discussion of a consistent picture of aging in the latter,
more complex case.

\section{General features of the reentrant system}

The $CdCr_{2-2x}In_{2x}S_4$ chromium thiospinel insulator \cite{spinel} has been
extensively characterized by neutron diffraction measurements
\cite{neutrons}.  For $x=0$, it is a ferromagnet ($T_c=84.5K$) with
ferromagnetic nearest neighbour interactions and antiferromagnetic
next-nearest neighbour interactions. For $x\ge 0.10$, the
ferromagnetic phase disappears, and a spin-glass phase appears at
$\sim 20K$ \cite{spinel,miyashita}. The $x=0.15$ system is a well
studied spin-glass compound \cite{Sitges,memchaos}. In the
intermediate region $0<x<0.10$, the ferromagnetic phase is followed at
lower temperature by a {\it reentrant} spin glass phase.

The main sample of our present study is the $x=0.05$ compound (of the
same batch as in \cite{neutrons}). Fig.1 shows its general magnetic
features.  The {\it dc} susceptibilities $\chi_{FC}$ and $\chi_{ZFC}$
have been measured along the usual field-cooled (FC) and zero-field
cooled (ZFC) procedures.  In the high temperature region, the {\it dc}
and {\it ac} susceptibilities follow a paramagnetic behaviour. At
$T_c\simeq 70K$, they rise up abruptly, reaching a plateau.  $\chi_{ZFC}$
and $\chi'$ at low frequency are at about the same level, $\sim 15\%$
below the $\chi_{FC}$ value.  The {\it ac} out-of-phase susceptibility
$\chi''$ peaks around $T_c$, and remains non-negligible in the
ferromagnetic plateau region, confirming the existence of magnetic
irreversibilities . At $T_g\sim 10K$, $\chi''$ shows a strong peak,
while $\chi'$ and $\chi_{ZFC}$ decrease with decreasing
temperature. In this low-temperature phase, spin glass features
coexist with those of a regular ferromagnetic state (non zero
magnetisation, Bragg peaks, spin waves \cite{spinel,neutrons}). The
{\it dc} plateau susceptibility in the ferromagnetic phase, which
corresponds approximately to the expected demagnetizing factor level,
indicates that the system (although disordered) is organized in
ferromagnetic domains.

\begin{figure}[htbp]
\begin{center}
   \leavevmode
   \epsfysize=5.6cm\epsfbox{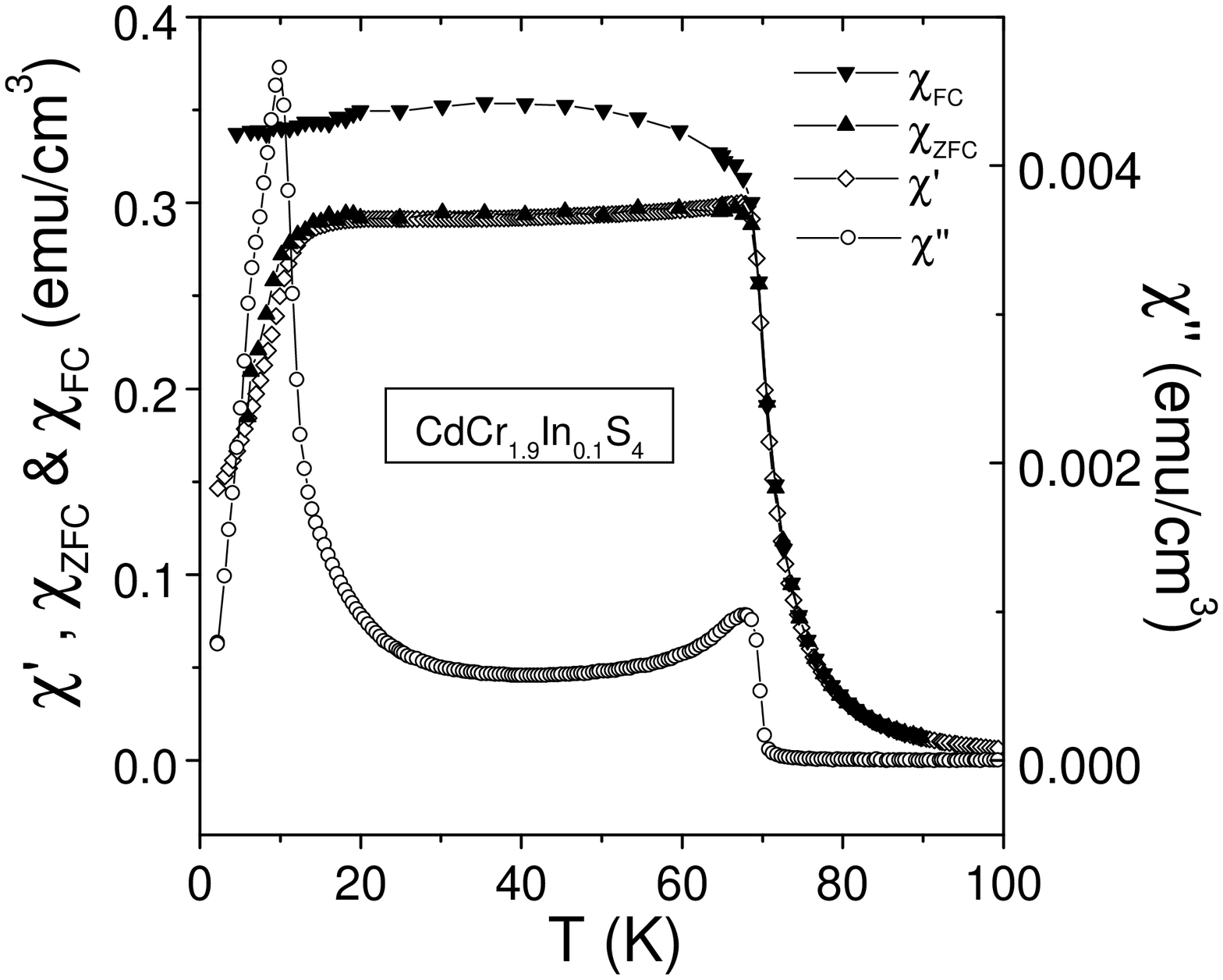}
    \hspace*{3mm}
    \epsfysize=5.6cm\epsfbox{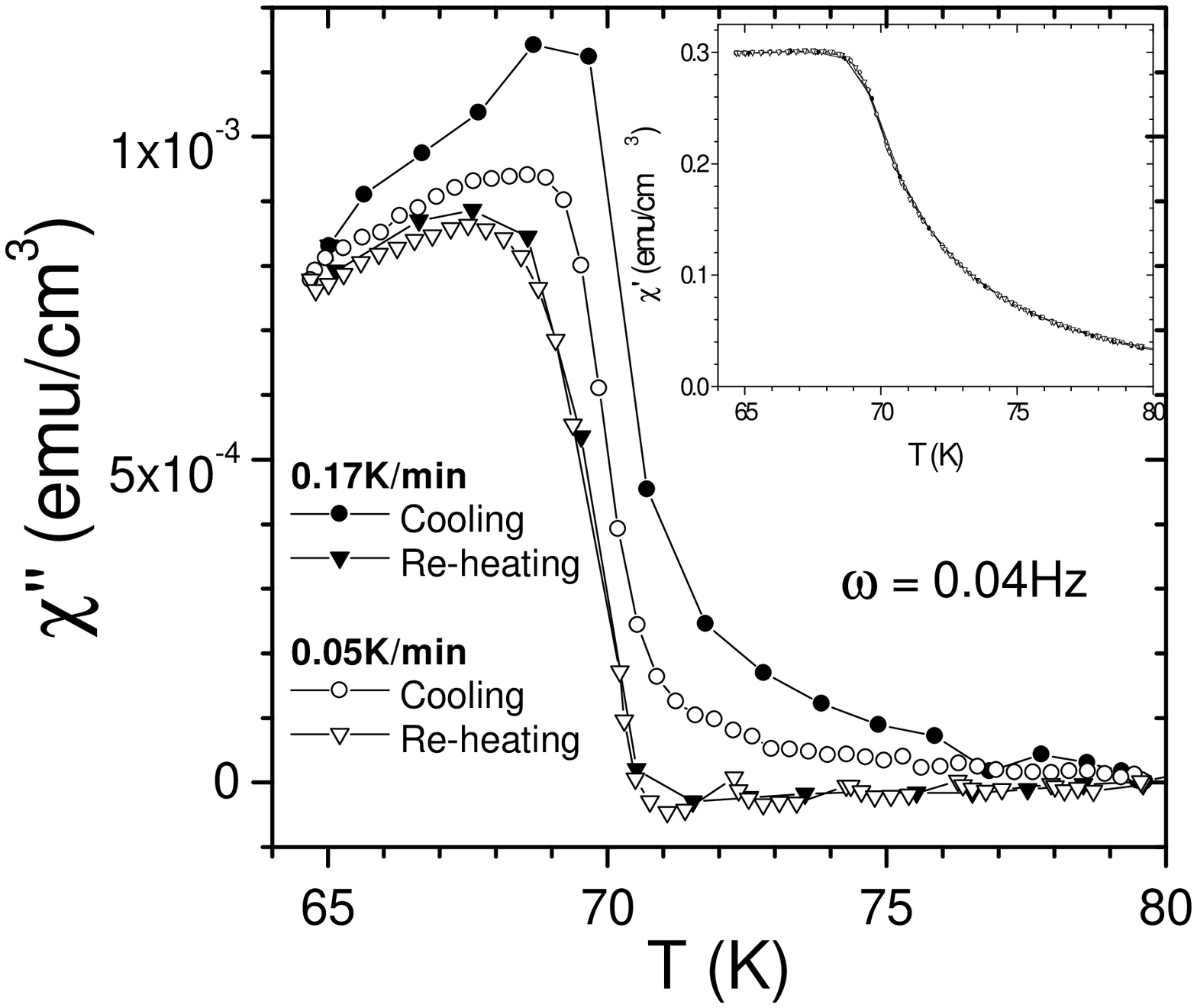}
  \end{center}
\vspace{.2cm}
  \hbox to \textwidth{\vspace*{5mm}\hfil Fig.\ \ref{ZFCFC}\hfil\hfil
  Fig.\ \ref{nagata}\hfil}

  \caption{\label{ZFCFC}
{\it ac} and {\it dc} suceptibilities of the $CdCr_{1.9}In_{0.1}S_4$ sample.
Both in-phase ($\chi'$) and out-of-phase ($\chi''$) components of
the {\it ac} susceptibility at frequency $0.04Hz$ are shown (open
symbols). The {\it dc} susceptibility (ZFC and FC, full symbols) has been measured
in a field of $10\ Oe$.
       }

  \caption{\label{nagata}
Hysteretic behaviour of $\chi''$ in the vicinity of the
ferromagnetic transition, upon cooling and re-heating at the
different rates of $0.17K/min$ (full symbols) or $0.05K/min$ (open
symbols). In this case, one can attribute the decrease of $\chi''$
to the decrease of the dissipative domain wall contribution. No
hysteresis is visible on $\chi'$ (inset).  }

\end{figure}

In Fig.2, we show $\chi_{ac}(T)$ measurements in the vicinity of the
ferromagnetic transition. The curves have been taken upon cooling from
the paramagnetic phase down to $65K$ and heating back, for two
different cooling/heating rates. $\chi''$ is clearly hysteretic (when
heating back it is always lower than during cooling), and sensitive to
the rate of temperature change (at a given temperature, $\chi''$ is
lower for slower rates).  This effect is much larger than in
spin-glasses, in which the cooling rate is to a large extent
irrelevant (see \cite{memchaos}). But it is similar to the behaviour
observed e.g. in a dipolar glass \cite{Levelut}, where thermally
activated domain growth is, as in the present case, the natural
scenario. In Fig.2, the obtained heating curve is the same for both
cooling/heating rates. The hysteresis seen
in Fig.2 is essentially determined during the time spent around $T_c$,
where domain walls are indeed expected to be weakly pinned.  Finally,
no significant hysteresis is seen on $\chi'$ (inset of Fig.2),
probably because $\chi'$, dominated by a volume response of the
domains, is much less sensitive than $\chi''$ to domain wall dynamics.

\section{Comparing aging phenomena in the spin-glass and
ferromagnetic phases}

We have applied to this system the procedure which allowed the
characterization of the so-called `memory and chaos effects' in spin glasses
\cite{memchaos}. 
The {\it ac} susceptibility has been measured at the three frequencies
of 0.04, 0.4 and 4 Hz in the same run.  We use the data in the
paramagnetic regime (assuming $\chi''=0$) for checking and
correcting slight frequency-dependent phase shifts in the
detection setup.

We first determine `reference curves' (solid lines in Fig.3-4): 
starting
from the paramagnetic phase, the temperature is continuously decreased
down to 3K at a rate of $0.1K/min$ , and then raised 
back to 100K at this same rate. 
 In a second 
experiment, we study aging in the following way. At 6 temperatures $T_i=67,
40, 20, 13, 8 \ \rm{and}\ 5K$, we have interrupted the cooling,
stabilized the temperature, and let the system age at constant
$T_i$ during 9 hours. When reaching 3K, we continuously re-heated
up to 100K (the whole measurement lasts about 3 days).

At each temperature $T_i$, $\chi''$ slowly relaxes downwards with
time.  In the {\it spin-glass} temperature region (Fig.3), the
$\chi''$ relaxation is the strongest in absolute as well as relative
value (amounting, for $0.04Hz$, resp. to $9,\ 20\ \rm{and}\ 10\%$ at
$5,\ 8,\ \rm{and}\ 13K$). The corresponding $\chi'$ relaxation,
although systematic, remains lower than $1\%$.  In the {\it
ferromagnetic} region (Fig.4), a $\chi''$ relaxation is also clearly
observed (although weaker than at low temperature, amounting for
$0.04Hz$ to resp. 4 and 5 \% at 40 and 67K).  Fig. 5 shows the
frequency dependence of the aging part of $\chi''$. In both
ferromagnetic and spin-glass phases, the same systematic trend of a
{\it stronger} relaxation at {\it lower} frequency is
found. Quantitatively, the curves can be rescaled onto a unique curve
as a function of the scaling variable $\omega(t+t_0)$ ($t_0$ being an
off-set time which takes into account the fact that the cooling
procedure is not instantaneous).  This scaling is equivalent to the
(approximate) $t/t_w$ scaling of {\it dc} experiments, which is
typically seen in spin-glasses (see discussions in
\cite{Sitges,Review}).

\begin{figure}[htbp]
\begin{center}
   \leavevmode
   \epsfysize=5.6cm\epsfbox{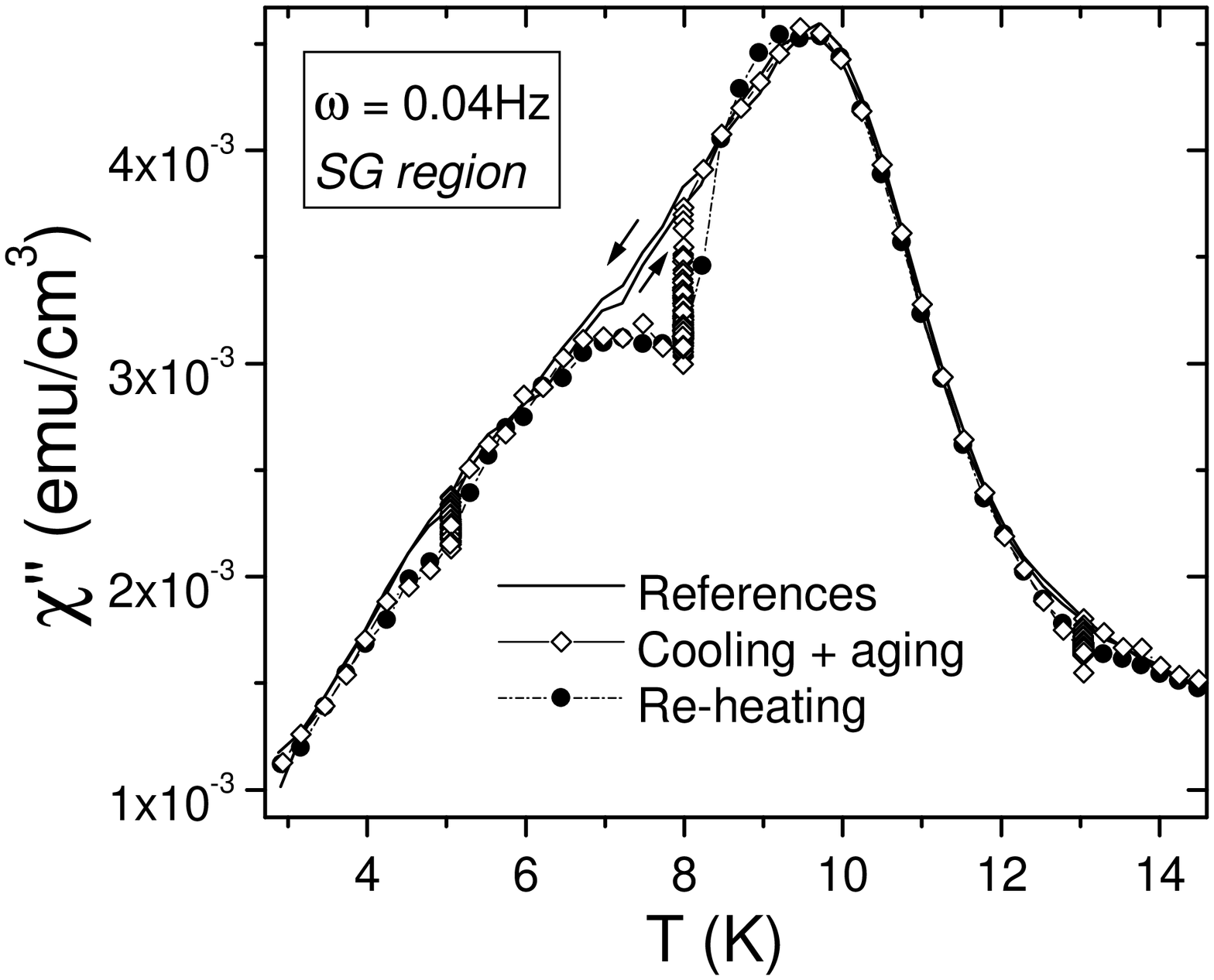}
    \hspace*{3mm}
    \epsfysize=5.6cm\epsfbox{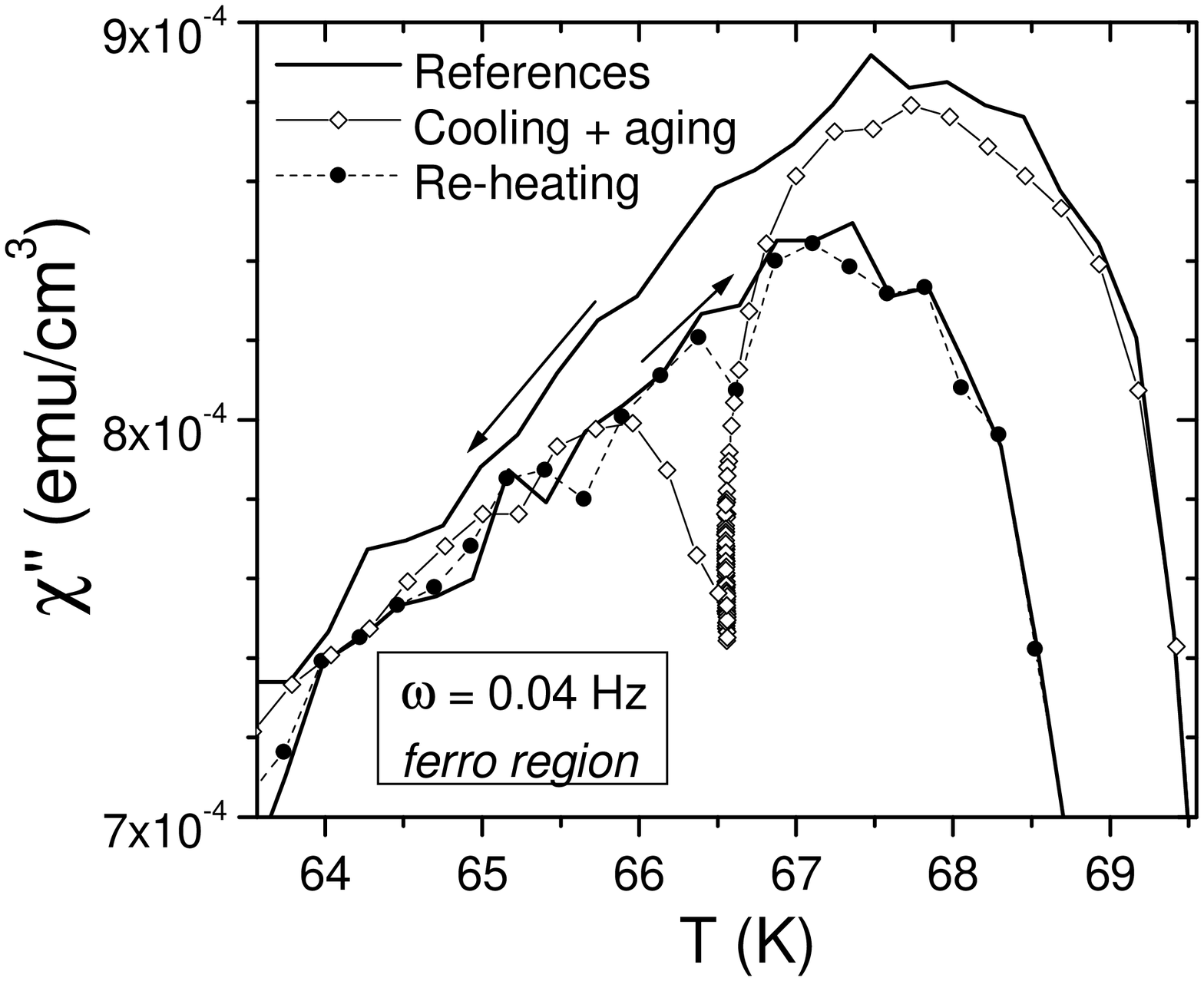}
  \end{center}
\vspace{.2cm}
  \hbox to \textwidth{\vspace*{5mm}\hfil Fig.\ \ref{relSG}\hfil\hfil
  Fig.\ \ref{relF}\hfil}

  \caption{\label{relSG}
Aging effects in the spin-glass region. The solid lines show a
reference behaviour for continuous cooling and re-heating (the
re-heating curve is slightly lower than the cooling curve). Open
diamonds: when cooling is stopped  at different temperatures
$T_i=13,\ 8\ {\rm and} \ 5K$ during 9h,  $\chi''$ relaxes due to
aging, but when cooling resumes  $\chi''$ merges with the
reference curve. Solid circles: when re-heating after cooling with
stops for aging, the memory of aging is retrieved.  }

  \caption{\label{relF}
Aging effects for $T_i=66.5K$ (beginning of the ferromagnetic
phase). Same experiment and conventions as in Fig.3, but here no
memory effect can be seen.}
\end{figure}

The most important point concerns the effect on aging of temperature
changes. At all temperatures $T_i$ (hence in both phases), we find
that when cooling is resumed after aging $\chi''$ merges back (even
increasing in some cases) with the reference curve, although this
reference is a {\it non-equilibrium} curve. This happens here exactly
like in spin glasses \cite{Sitges,hierarki,DTupps,memchaos}, in a
similar fashion as in some other systems \cite{KTN, ciliberto}.  Aging
processes have to restart at lower temperatures, as if the aging
evolution was not {\it cumulative} with that at a higher temperature
(`chaos-like' or `rejuvenation' effect), in contrast with the
continuous hysteresis phenomenon displayed in Fig.2.  Another
important feature is seen in the $\chi''(T)$ curve taken upon
re-heating steadily from $3K$. When reaching each of the temperatures
$T_i$ in the spin-glass region (5, 8 and 13K), $\chi''(T)$ departs
from the reference curve and traces back a dip which displays the {\it
memory} of the past relaxation at $T_i$.  Such memory effects have
been characterized in details in spin glasses
\cite{Sitges,hierarki,DTupps,memchaos}. In the {\it ferromagnetic}
phase (20, 40 and 67K), in contrast, no memory effect is found with
this procedure (Fig. 4).  This result is in agreement with previous
{\it ac} and {\it dc} temperature cycling experiments performed on
another reentrant system \cite{ferroupps}, in which however disorder
effects seem to be stronger than here (larger FC-ZFC separation,
widely rounded ferromagnetic $\chi_{FC}$ plateau, a behaviour which is
very similar to that of the more diluted $x=0.90$ thiospinel
\cite{spinel}).  In \cite{ferroupps}, it is observed that the effect
of {\it positive} and {\it negative} temperature cycles is similar in
the ferromagnetic phase (rejuvenation-type upon cooling {\it as well
as} heating), while it is strongly asymmetric in the spin-glass phase
(rejuvenation upon cooling, memory upon heating). Hence, rejuvenation
effects are found in these disordered ferromagnets like in spin
glasses, but memory effects are not readily seen in the ferromagnetic
phase, where domain growth processes are important. In the following,
we discuss wall dynamics as a possible source of interplay between
spin-glass like and domain growth processes, and show that memory
effects can be obtained in the ferromagnetic phase provided that the
low-temperature cycling is short enough.

\begin{figure}[htbp]
\begin{center}
   \leavevmode
   \epsfysize=5.4cm\epsfbox{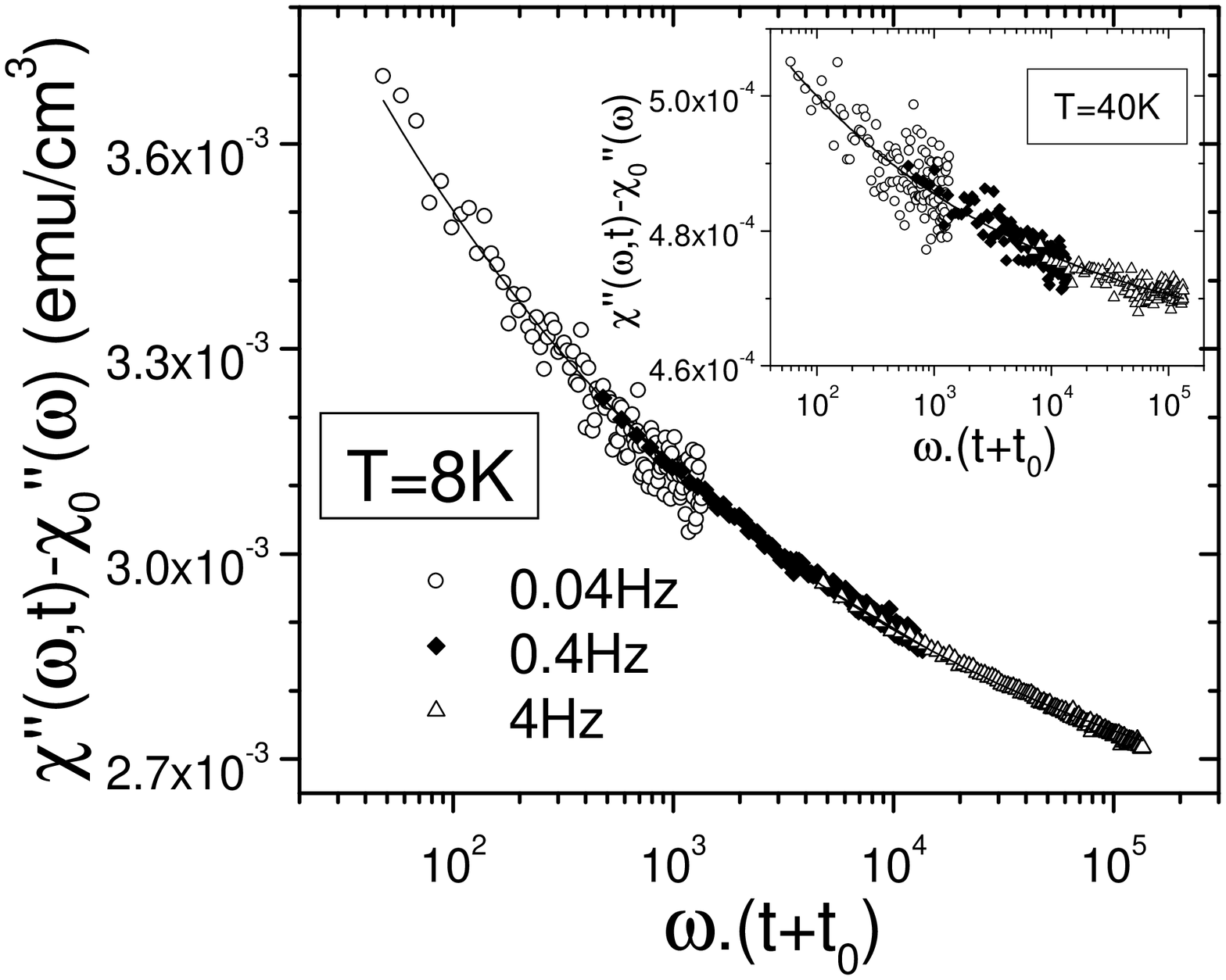}
    \hspace*{3mm}
    \epsfysize=5.4cm\epsfbox{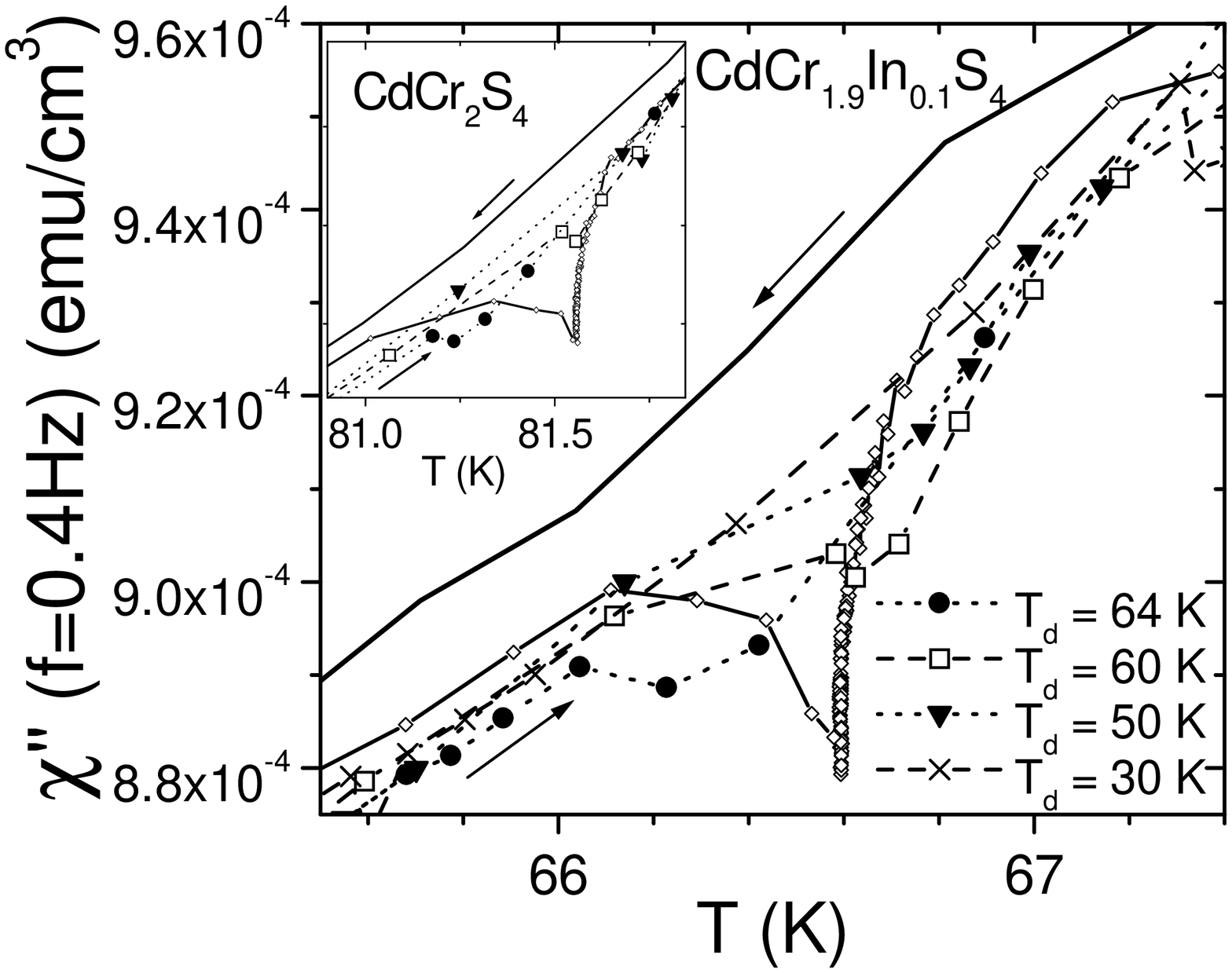}
  \end{center}
\vspace{.2cm}
  \hbox to \textwidth{\vspace*{5mm}\hfil Fig.\ \ref{fig5}\hfil\hfil
  Fig.\ \ref{fig6}\hfil}

  \caption{\label{fig5}
Scaling of the $\chi''$-relaxation curves at $T_i=8K$ and $T_i=40K$
for different frequencies $0.04,\ 0.4\ {\rm and}\ 4Hz$. The
curves have been shifted vertically in order to compensate for the
different equilibrium values, and an off-set time $t_0$ has been added
to account for the fact that the cooling procedure is not
instantaneous.  We have also shown for comparison the fit proposed in
\protect\cite{BD,Review}:
$\chi''(\omega,t)=\chi''_{eq}(\omega)+[\omega(t+t_0)]^{-b}$, with
$b=0.2$ for both temperatures.}

  \caption{\label{fig6}
Partial memory effect in the ferromagnetic phase. The procedure is similar to that of Fig.3-4,
except that the low-temperature excursion was limited to either
$T_d=64K$ (full circles), $60K$ (open squares), $50K$ (full triangles) or $30K$
(crosses). For a short enough low-temperature history (e.g.
$T_d=64K$, full circles), the re-heating curves show a partial memory
effect, which progressively faints out when significant
ferromagnetic domain growth processes can occur (i.e. for
decreasing $T_d$ values). The main figure concerns the
$CdCr_{1.9}In_{0.1}S_4$ compound; in the inset, we show the same
effect on the non-diluted $CdCr_{2}S_4$ ferromagnet (same conventions
as in main figure, with $T_d$= $78,\
75,\ 65K$).
}

\end{figure}

\section{Ferromagnetic aging: a combination of domain growth and
spin-glass like dynamics}

In an ideal ferromagnet, domain growth only involves microscopic time
 scales.  In the presence of pinning disorder (here dilution of $Cr$
 by $In$, or structural defects), a domain wall has many metastable
 configurations between which it makes thermally activated hops. This
 gives rise to slow dynamics at the laboratory time scale.  Actually,
 the problem of pinned elastic interfaces has similarities with that
 of spin-glasses. The frustration comes from the elastic energy
 associated to a deformation of the wall, which limits the possible
 excursions between the favourable pinning sites. Several theoretical
 arguments \cite{DWhierarchy} suggest that the energy landscape of a
 pinned domain wall is {\it hierarchical}, with small length scale
 $\ell$ reconformations corresponding to small energy barriers
 $E(\ell)\sim \Upsilon \ell^\theta$.

In the spin glass phase, the {\it rejuvenation} upon cooling and the
{\it memory} when heating back have been ascribed to a hierarchical
organization of the metastable states, which are progressively
uncovered as the temperature is decreased \cite{hierarki,Sitges,BD}
(see also \cite{Cugliandolo}). In brief, when the temperature is
reduced, the system remains in a deep well (allowing for the `memory
effect'), while new subwells appear, inducing new aging processes
(rejuvenation effect).  For a pinned domain wall, the hierarchy of
reconformation length scales \cite{DWhierarchy} is an appealing
transposition in `real space' of the spin-glass hierarchy
of metastable states, in the following sense.  As time elapses at
fixed temperature, wall reconformations occur at longer and longer
length scales (aging), while shorter length processes $\ell < \ell^*$,
with $E(\ell^*)=kT$, are fluctuating in equilibrium. When the
temperature is lowered, the long wavelength modes become frozen (thus
corresponding to the deep well, which allows the memory), and the
`glass length' $\ell^*$ decreases to $\ell^{*\prime}$. The modes such
that $\ell^{*\prime} < \ell < \ell^*$, on the other hand, are no longer
in equilibrium because the Boltzman weights of the different
`sub-wells' have changed, and start aging (rejuvenation)
\cite{BD,DWhierarchy}. These reconformations are expected to
contribute to $\chi''$ as a function of $\omega t$
\cite{DWhierarchy,Yoshino}, as indeed observed here in both phases.

In the ferromagnetic phase (and particularly close to $T_c$), {\it in
parallel} with these reconformation processes, the average domain size
grows with time.  Thus, the impurities with which a domain wall
interacts are changing.  Obviously, this {\it net motion} of the
domain walls cannot preserve the memory of the reconformations.
Domain growth effects are directly visible in the overall decrease of
$\chi''$ between cooling and heating, a strong effect close to the
transition at 70K (Fig.2 and 4), also present (slighter, but
systematic) in the spin-glass phase (Fig.3). The absence of memory in
Fig.4 is observed after a long period spent at lower temperatures,
during which 
it is likely that the positions of the domain walls have indeed
significantly changed.

We have therefore performed a slightly different experiment, with the
aim of controlling the effect of domain growth after the aging
relaxation.  After aging $3.5h$ at $T_i=66.6K$, the sample is cooled
to a temperature $T_d$ which remains close to $T_i$, and then
re-heated. In order to still accelerate the procedure, we have
suppressed the measurement at the lowest frequency of $0.04$Hz.  Fig.6
shows that when the low-temperature excursion is small enough (like
$T_d=64K$), a partial memory of aging is revealed.  The curves are
somewhat noisy, but the effect is systematic as a function of $T_d$ :
for lower and lower $T_d$ values (60, 50 and 30K), the dip
progressively disappears \cite{Levelutpriv}. The same experiment, performed in the middle
of the ferromagnetic plateau ($T_i=40K$, $T_d=30\ {\rm and}\ 35K$),
shows the same qualitative trend (although with weaker relaxation)
\cite{VD}. Furthermore, we have recently started the investigation of
the {\it non-diluted} $CdCr_2S_4$ ferromagnet
\cite{spinel,neutrons}. The first results (inset of Fig.6) do again
show the same `rejuvenation and partial memory' effects. In this
non-diluted system, there are also competing interactions
\cite{spinel}. Further studies on ferromagnets with non-frustrated
interactions are needed in order to clarify the respective
contributions of the frustration of interactions on the one hand, and
of pinning mechanisms on defects of various origins on the other hand
\cite{VD}. 

Finally, let us note that a similar systematic
shift of the dip position with $T_d$ is observed in both samples of
Fig.6.  This is reminiscent of the way in which the memory is erased
by a slight re-heating in the $x=0.85$ thiospinel spin glass (see 2nd
ref. of \cite{memchaos}), and is yet unexplained.

\section{Conclusions}

Within the domain-growth description of spin-glasses \cite{domains},
the way to interpret the rejuvenation effect (and the weak cooling
rate dependence of the measured quantities) is to invoke the idea of
the chaotic evolution of the spin-glass order with temperature
\cite{bray}. However, as emphasized in \cite{memchaos}, this is
difficult to reconcile with the simultaneous memory effect discussed
above. Furthermore, such chaos should be absent in the ferromagnetic
phase (or indeed only concern the domain wall conformations). Let us
finally note that until now numerical simulations have failed in identifying 
this type of `chaos' \cite{numrecent}.

On the other hand, the coexistence of memory and rejuvenation has
 readily been interpreted in a hierarchical phase space picture
 \cite{hierarki,BD}, which finds in the present case a natural real
 space interpretation in terms of wall reconformations
 \cite{DWhierarchy}.  Comparing the spin-glass and ferromagnetic
 situations suggests that the dynamics in spin glasses is dominated by
 {\it wall reconformation} processes and not by {\it domain growth},
 which would erase the memory, and lead to strong cooling rate
 dependence.  The extrapolation of this idea to the case of spin
 glasses raises incentive questions about the nature of `domains' and
 `walls' in this case. Many ideas on the non-trivial geometry of the
 spin-glass domains have indeed been proposed over the years
 \cite{fractdiv,BD,memchaos,martin}. A possibility is that the ordered
 phase of a spin-glass contains a large number of pinned, zero tension
 walls, which reconform in their disordered landscape without any
 overall tendency to coarsen. The growth of a spin-glass correlation
 length, reported both numerically \cite{numerics} and experimentally
 \cite{orbachbar}, could then be understood as the progressive
 increase of a typical length scale for wall reconformations at fixed
 temperature, while the effect of temperature variations would involve
 a hierarchy of different length scales.

\stars

\section{Acknowledgements}
We are grateful to F. Alet for his active participation to this
work. We benefited from stimulating discussions with M. Ocio, J.
Houdayer, A. Levelut, O. Martin, S. Miyashita and H. Yoshino, and thank L. Le
Pape for his skillfull technical support.

\end{document}